\documentclass[twocolumn,prb,superscriptaddress]{revtex4-1}
\usepackage[latin1]{inputenc}
\usepackage{graphicx}
\usepackage{amssymb}
\usepackage{amsmath}
\usepackage{xspace}
\usepackage{dcolumn}
\usepackage{bm}
\usepackage{color}
\usepackage[extra]{tipa}

\newfont{\tensy}{cmsy10}

\newcommand{\ie}[0]{i.e.\@\xspace}

\newcommand{\rmd}{\text{d}}
\newcommand{\rmi}{\text{i}}
\newcommand{\UP}[0]{\uparrow}
\newcommand{\DO}[0]{\downarrow}

\newcommand{\oP}{\hat{P}}
\newcommand{\oQ}{\hat{Q}}

\newcommand{\on}{\hat{n}}

\newcommand{\kB}{k_\text{B}}
\newcommand{\nag}{{\phantom{\dag}}}

\newcommand{\las}[0]{\langle}
\newcommand{\ras}[0]{\rangle}

\newcommand{\RefsAB}{\cite{ohgoe2017competitions,1709.00278}\xspace}

\setcitestyle{square,numbers}

\begin{document}


\title{Dominant charge-density-wave correlations in the Holstein model\\ on the half-filled
  square lattice}

\author{M. Hohenadler}

\affiliation{\mbox{Institut f\"ur Theoretische Physik und Astrophysik,
    Universit\"at W\"urzburg, 97074 W\"urzburg, Germany}}

\author{G. G. Batrouni}

\affiliation{Universit\'e C\^ote d'Azur, CNRS, INPHYNI, France}
\affiliation{MajuLab, CNRS-UCA-SU-NUS-NTU International Joint Research
  Unit, 117542 Singapore}
\affiliation{Centre for Quantum Technologies, National University of
  Singapore, 2 Science Drive 3, 117542 Singapore}
\affiliation{Department of Physics, National University of Singapore, 2
  Science Drive 3, 117542 Singapore}
\affiliation{Beijing Computational Science Research Center, Beijing 100193, China}

\begin{abstract}
We use an unbiased, continuous-time quantum Monte Carlo method to
address the possibility of a zero-temperature phase without
charge-density-wave (CDW) order in the Holstein and, by extension,
the Holstein-Hubbard model on the half-filled square lattice.  In particular, we present results
spanning the whole range of phonon frequencies, allowing us to use the well
understood adiabatic and antiadiabatic limits as reference points.
For all parameters considered, our data suggest that CDW correlations
are stronger than pairing correlations even at very low temperatures. These
findings are compatible with a CDW ground state that is also suggested by theoretical arguments.
\end{abstract}

\date{\today}

\maketitle


\section{Introduction}\label{sec:introduction}

Charge-density-wave (CDW) and superconducting (SC) phases are ubiquitous in
quasi-two-dimensional (quasi-2D) materials and often arise from
electron-phonon coupling. Holstein's molecular-crystal model \cite{Ho59a} of electrons
coupled to quantum phonons has played a central role for the investigation of
such phenomena. However, even after decades of research, fundamental
questions are still unanswered. 
Quantum Monte Carlo (QMC) approaches have played a key role
in the study of this problem. Despite important recent methodological
advances \cite{PhysRevB.98.085405,arXiv:1704.07913,BaSc2018,PhysRevB.98.041102,Ka.Se.So.18,li2019accelerating},
simulations of electron-phonon problems remain significantly more challenging
than, for example, those of purely fermionic Hubbard models.

\begin{figure}
\includegraphics[width=0.425\textwidth]{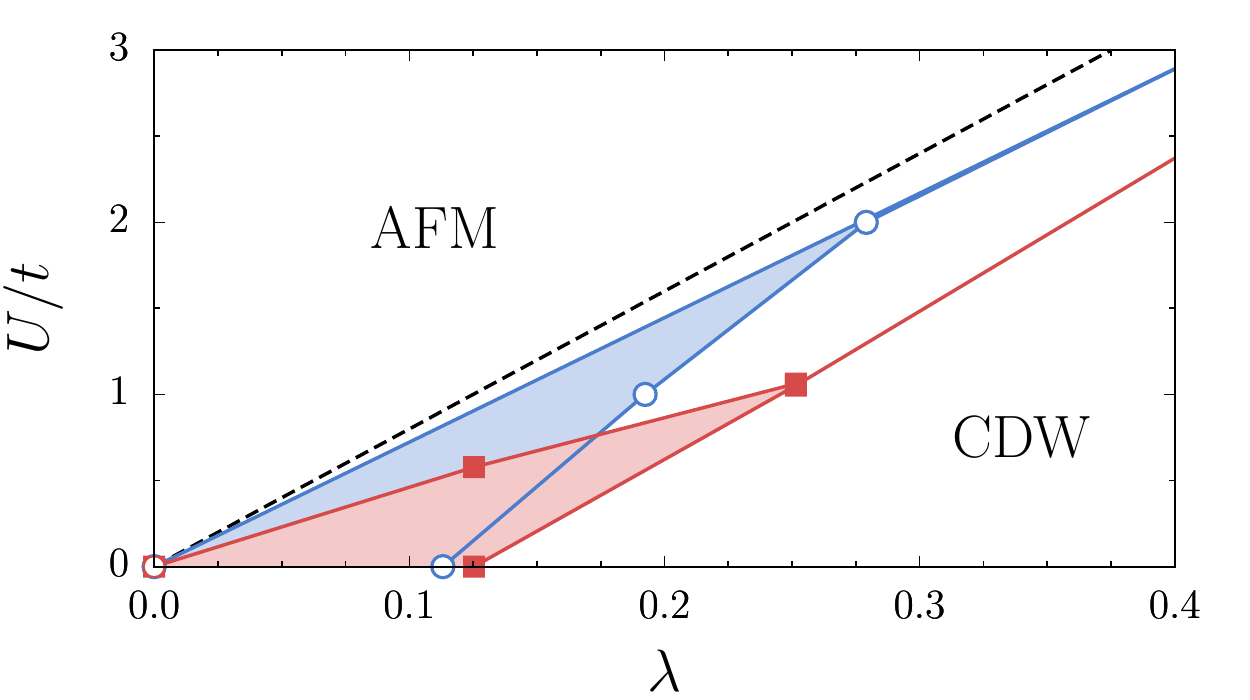}
\caption{\label{fig:phasediagrams}
  Ground-state phase diagram of the Holstein-Hubbard model on a half-filled
  square lattice for $\omega_0/t=1$, as suggested by variational QMC
  calculations in Ref.~\cite{ohgoe2017competitions} (open symbols) and Ref.~\cite{1709.00278} (filled symbols). The shaded regions
  were reported to be either metallic or superconducting, without long-range
  CDW or AFM order. The dashed line corresponds to $U=\lambda W$ with $W=8t$.
  The definition of $\lambda$ in Ref.~\cite{1709.00278} differs from ours and
  Ref.~\cite{ohgoe2017competitions} by a factor of $8$. Here,
  we focus on the Holstein limit $U=0$.
}
\end{figure}

The Holstein-Hubbard model captures the interplay of electron-phonon coupling ($\sim\lambda$)
and electron-electron repulsion ($\sim U$). For $U=0$, it reduces to
the Holstein model simulated in the following. For the much studied case of a half-filled
square lattice, earlier work \cite{PhysRevB.52.4806,PhysRevLett.75.2570,PhysRevB.75.014503,PhysRevB.92.195102,PhysRevLett.109.246404,PhysRevB.87.235133,ohgoe2017competitions}
agreed on either long-range CDW or AFM order at $T=0$ depending on $\lambda$ and
$U$, consistent with theoretically expected instabilities of the Fermi liquid.
In contrast, relying on variational QMC simulations, two recent papers
\RefsAB reported the existence of an intermediate metallic phase with neither
CDW nor AFM order, see Fig.~\ref{fig:phasediagrams}. Instead, it was
characterized as either SC or paramagnetic \RefsAB. 
The prospect of a metallic ground state has to be distinguished from metallic
behavior emerging at finite temperatures \cite{PhysRevB.87.235133} simply via
the thermal destruction of AFM order \cite{PhysRevB.98.085405}.
The predicted existence of the metallic state even at $U=0$ (see
Fig.~\ref{fig:phasediagrams}), \ie, the Holstein model, appears to rule
out competing interactions as its origin.

Apart from the challenges due to small gaps and order
parameters at weak coupling, the methods used in Refs.~\RefsAB are
variational and involve an ansatz for the ground-state wave function that may bias the results. A similar
controversy regarding metallic behavior in the 1D
Holstein-Hubbard model was recently resolved. For the latter, approximate
strong-coupling results in combination with unfounded conclusions from
numerical simulations \cite{Hirsch83a} as well as insufficiently accurate
renormalization-group (RG) approaches were contradicted by unbiased numerical
simulations and functional RG calculations. For the 1D case, a disordered
phase has been firmly established \cite{MHHF2017}, although claims of dominant pairing correlations \cite{ClHa05} in this regime
were refuted \cite{PhysRevB.92.245132}. Whereas the Fermi liquid is expected to
be unstable at $T=0$ in the particular 2D setting considered, Refs.~\RefsAB
also suggest that the non-CDW region
could have SC order. SC correlations have been found to be enhanced in Holstein models with
next-nearest-neighbor hopping \cite{PhysRevB.46.271}, dispersive phonons
\cite{PhysRevLett.120.187003}, frustration
\cite{li2018superconductivity}, or finite doping \cite{PhysRevB.40.197}. 
An extended semimetallic phase is supported by theory and
numerics in the Holstein model on the honeycomb lattice \cite{PhysRevLett.122.077601,PhysRevLett.122.077602}.

Here, to provide further insight into this problem in the limit $U=0$,
we exploit the properties of
the continuous-time interaction-expansion (CT-INT) QMC method \cite{Rubtsov05}. Compared to other
approaches, it can in principle access rather low temperatures in the
weak-coupling regime. Although simulations are partially
restricted by a sign problem, we obtain evidence for long-range CDW order
at very low temperatures for parameters where a non-CDW phase was predicted \RefsAB.

The paper is organized as follows. In Sec.~\ref{sec:model}, we define the
model and summarize previous work and theoretical
arguments. Section~\ref{sec:method} provides the necessary details about the
CT-INT simulations. Our results are discussed in Sec.~\ref{sec:results},
followed by our conclusions in Sec.~\ref{sec:conclusions}.

\section{Model}\label{sec:model}

References~\cite{ohgoe2017competitions,1709.00278} presented phase diagrams
for the Holstein-Hubbard model on the half-filled square
lattice. Selected results are reproduced in Fig.~\ref{fig:phasediagrams}.
Because the purported non-CDW region is most extended for a vanishing Hubbard
repulsion ($U=0$), we focus on the simpler Holstein Hamiltonian \cite{Ho59a}
\begin{align}\label{eq:model}
  \hat{H}
  =
    -t \sum_{\las i,j\ras \sigma} \hat{c}^\dag_{i\sigma} \hat{c}^\nag_{j\sigma} 
    +
    \sum_{i}
    \left[
    \mbox{$\frac{1}{2M}$} \oP^2_{i}
    +
    \mbox{$\frac{K}{2}$} \oQ_{i}^2
    \right]
    -
      g
    \sum_{i} \hat{Q}_{i} 
     \hat{\rho}_i   
    \,.
\end{align}
Here, $\hat{c}^\dag_{i\sigma}$ creates an electron with spin $\sigma$ at lattice
site $i$ and the first term describes nearest-neighbor hopping with amplitude
$t$. Lattice vibrations are modeled in terms of independent harmonic oscillators with frequency
$\omega_0=\sqrt{K/M}$, displacements $\hat{Q}_i$, and momenta
$\hat{P}_i$. The electron-phonon interaction takes the form of a
density-displacement coupling to local fluctuations of the electron number;
we have
$\hat{\rho}_i=\on_i-1$ with $\on_{i} =
\sum_\sigma \on_{i\sigma}$ and $\on_{i\sigma} = \hat{c}^\dag_{i\sigma}\hat{c}^\nag_{i\sigma}$.
All simulations were done on $L\times L$ square lattices with periodic
boundary conditions and for a half-filled band  ($\las\on_{i}\ras=1$,
chemical potential $\mu=0$). We define a dimensionless coupling constant
$\lambda=g^2/(W K)$ ($W=8t$ is the free bandwidth), set $\hbar$, $\kB$, and
the lattice constant to one and use $t$ as the energy unit.

Hamiltonian~(\ref{eq:model}) has been the subject of numerous QMC
investigations \cite{PhysRevB.40.197,PhysRevB.42.2416,PhysRevB.42.4143,PhysRevLett.66.778,PhysRevB.43.10413,PhysRevB.46.271,PhysRevB.48.7643,PhysRevB.48.16011,PhysRevB.55.3803}.
With the exception of Refs.~\RefsAB, a CDW ground state 
was assumed to exist for any $\lambda>0$, as suggested by
several theoretical arguments. First, for classical phonons, corresponding to
$\omega_0=0$, mean-field theory is exact at $T=0$ and reveals a gap and CDW order for any
$\lambda>0$ \cite{PhysRevLett.66.778}. The origin of this weak-coupling instability is the combination of perfect nesting on a half-filled
square lattice with nearest-neighbor hopping and a zero-energy Van Hove
singularity in the density of states \cite{PhysRevLett.56.2732,PhysRevB.42.2416}. These features give rise to a 
noninteracting charge susceptibility [defined in Eq.~(\ref{eq:chi}) below] that
diverges as $\chi^{(0)}_\text{CDW}\sim\ln^2\beta t$
\cite{PhysRevLett.56.2732,PhysRevB.42.2416}, where $\beta=1/T$. 
Both, at the mean-field level and in numerical simulations,
such a divergence produces CDW order
for any $U<0$ in the attractive Hubbard model  \cite{Hirsch85}. The latter is
an exact limit of the Holstein model for $\omega_0\to\infty$ ($M\to
0$), with $U=\lambda W$. Hence, long-range CDW order at $T=0$ is established
for the Holstein model both for $\omega_0=0$ and $\omega_0=\infty$. In contrast,
the s-wave pairing susceptibility [Eq.~(\ref{eq:chip})] has a weaker divergence,
$\chi^{(0)}_\text{SC}\sim\ln\beta t$, because nesting plays no role.
This is consistent with the observation that for $\omega_0<\infty$
SC correlations are weaker than CDW correlations at half-filling
\cite{PhysRevLett.66.778,PhysRevB.40.197} but not with an SC phase~\RefsAB.
However, earlier work did not consider the weak-coupling regime and
unbiased, high-precision finite-size scaling analyses only appeared recently
\cite{PhysRevB.98.085405,PhysRevB.98.041102,BaSc2018,li2019accelerating}.

In Refs.~\RefsAB, the challenging problem of determining the ground-state
phase diagram was approached using zero-temperature variational QMC methods.
In contrast, most other work (for exceptions see
Refs.~\cite{PhysRevB.52.4806,li2018superconductivity}) infer ground-state properties
from simulations at low but finite temperatures. Whereas the AFM phase of the
Holstein-Hubbard model shown in Fig.~\ref{fig:phasediagrams} exists only at
$T=0$, long-range CDW order is associated with an Ising order parameter and persists
up to a critical temperature $T^\text{CDW}_c$
\cite{PhysRevLett.66.778,PhysRevB.98.085405}. Similarly, the U(1) SC
order parameter also permits a nonzero transition temperature $T^\text{SC}_c$.
In both cases, given an ordered ground state, we therefore expect a
finite-temperature phase transition.  An important
exception is the limit $\omega_0=\infty$, corresponding to the attractive Hubbard
model. The latter has an enhanced symmetry that combines the CDW and SC order
parameters into an SU(2) vector \cite{Hirsch85}. According to the Mermin-Wagner
theorem \cite{PhysRevLett.17.1133}, long-range order is therefore confined to $T=0$.

Let us address the purported intermediate phase at weak
coupling and $\omega_0>0$ reported in Refs.~\RefsAB in the light of these arguments.
The overall size of this phase increases with increasing $\omega_0/t$ in
Refs.~\RefsAB, similar to the case of the 1D Holstein-Hubbard model \cite{MHHF2017}.
In 1D, quantum lattice fluctuations promote the proliferation of domain walls in the
Ising CDW order parameter. There, the ground state is metallic up to
$\lambda=\lambda_c$ with $\lambda_c\to\infty$ for  $\omega_0\to\infty$
(attractive Hubbard model) \cite{MHHF2017}.
In contrast, the 2D Holstein model is CDW-ordered in the antiadiabatic limit $\omega_0\to\infty$.
An explicit comparison of data for $\omega_0=\infty$ and $\omega_0<\infty$
will be made in Sec.~\ref{sec:results}. No theoretical arguments
were given in Refs.~\RefsAB against the weak-coupling instability 
expected from the divergence of $\chi^{(0)}_\text{CDW}$.
Interestingly, although $\chi_\text{CDW}^{(0)}$ and  $\chi_\text{AFM}^{(0)}$
diverge in the same way, the methods of Refs.~\RefsAB successfully detect the
weak-coupling AFM instability at $\lambda=0$ but not the CDW instability at
$U=0$ (see Fig.~\ref{fig:phasediagrams}).
Another apparent inconsistency is that the non-CDW region at $U=0$ is
significantly larger for $\omega_0/t=8$ than for $\omega_0/t=1$ in
Ref.~\cite{ohgoe2017competitions}, whereas
it remains virtually unchanged between $\omega_0/t=1$ and $\omega_0/t=15$ in
Ref.~\cite{1709.00278}. For the value $\omega_0/t=1$ analyzed in both works
and shown in Fig.~\ref{fig:phasediagrams}, Ref.~\cite{ohgoe2017competitions}
predicts a non-CDW ground state up to $\lambda\approx 0.11$ at $U=0$, whereas
Ref.~\cite{1709.00278} reports a critical value of $\lambda\approx 0.125$
(using our definition of $\lambda$).

\section{Method}\label{sec:method}

The application of the CT-INT method \cite{Rubtsov05} to electron-phonon
models goes back to the work by Assaad and Lang \cite{Assaad07}. For
investigations of 2D  Holstein and Holstein-Hubbard models, see
Refs.~\cite{PhysRevB.98.085405,PhysRevLett.122.077601}.
Its general,
action-based formulation makes it suitable for retarded fermion-fermion
interactions that arise naturally from electron-phonon problems after
integrating out the phonons in the path-integral representation of the
partition function \cite{Assaad07}. The weak-coupling expansion can be shown to converge for
fermionic systems in a finite space-time volume \cite{Rubtsov05}, so that the
method is exact apart from statistical errors. General reviews
have been given in Refs.~\cite{Gull_rev,Assaad14_rev}. 

The numerical effort scales cubically with the average expansion order $n$,
where $n\approx {O}(\beta\lambda L^2)$ for the Holstein model.
While other methods formally scale linearly in $\beta$ \cite{Assaad18_rev}, CT-INT is typically less
limited by autocorrelation times. For the present work, its use is motivated by
a significant speedup at small $\lambda$ that permits us to study reasonably
large system sizes up to $L\leq12$ at inverse temperatures $\beta t\leq 96$. For
intermediate phonon frequencies, the method is ultimately limited by a sign
problem that arises from the absence of an exact symmetry between the
spin-$\UP$ and spin-$\DO$ sectors \cite{PhysRevB.98.085405}. The results for $\omega_0=\infty$
were obtained by directly simulating the attractive Hubbard model with the CT-INT method.
We used 1000 single-vertex updates and 8 Ising spin flips per
sweep for all simulations. Although our method is entirely unbiased, as
opposed to the algorithms of Refs.~\RefsAB, limitations arise regarding
model parameters, temperatures, and system sizes.

\section{Results}\label{sec:results}

To detect CDW and/or s-wave SC order, we carried out a finite-size scaling analysis
based on the charge and pairing susceptibilities (with $\hat{\Delta}_i=\hat{c}_{i\UP}\hat{c}_{i\DO}$)
\begin{align}\label{eq:chi}
  \chi_\text{c}(\bm{q})  
  &=
    \frac{1}{L^2} \sum_{ij} e^{\rmi(\bm{r}_i-\bm{r}_j)\cdot\bm{q}} \int_0^\beta \rmd \tau
  \las \hat{n}_{i}(\tau) \hat{n}_{j}\ras\,,
  \\\label{eq:chip}
  \chi_\text{p}(\bm{q})    
  &= \frac{2}{L^2} \sum_{ij}
  e^{\rmi(\bm{r}_i-\bm{r}_j)\cdot\bm{q}} \int_0^\beta \rmd \tau
  \las \hat{\Delta}^\dag_{i}(\tau) \hat{\Delta}^\nag_{j}\ras\,,
\end{align}
We define $\chi_\text{CDW}\equiv\chi_\text{c}(\bm{Q}_\text{CDW})$ with
$\bm{Q}_\text{CDW} = (\pi,\pi)$ and
$\chi_\text{SC}\equiv\chi_\text{p}(\bm{Q}_\text{SC})$ with $\bm{Q}_\text{SC}= (0,0)$.
The factor $2$ in the definition of $\chi_\text{p}$ ensures
$\chi_\text{CDW}\equiv\chi_\text{SC}$ for the attractive Hubbard model
($\omega_0\to\infty$) and at $\lambda=0$.

Long-range CDW order can be detected from by the renormalization-group
invariant correlation ratio
\begin{align}\label{eq:Rchic}
  R^\chi_\text{CDW} 
  &=  1-\frac{\chi_\text{c}(\bm{Q}_\text{CDW}-\delta{\bm
    q})}{\chi_\text{c}(\bm{Q}_\text{CDW})}\,,\quad |{\bm q}|=\frac{2\pi}{L}\,.
\end{align}
At a fixed $\lambda$, $R^\chi_\text{CDW}$ depends only on $L^z/(T-T^\text{CDW}_c)$, so
that data for different $L$ are expected to intersect (up to corrections to
scaling) at the transition temperature $T^\text{CDW}_c$. By definition,
$R^\chi_\text{CDW}\to 0$ as $L\to\infty$ in the absence of long-range CDW
order, whereas $R^\chi_\text{CDW}\to 1$ as $L\to\infty$ if
$\chi_\text{CDW}$ diverges with $L$. The ratio $R^\chi_\text{CDW}$ can be
expected to have smaller scaling corrections than $\chi_\text{CDW}$ itself
\cite{Binder1981}.  The use of susceptibilities rather than static
structure factors suppresses background contributions to critical fluctuations.
We will also consider the finite-size scaling of the susceptibility itself,
which is described near the Ising critical point by the scaling form
\begin{equation}
  \chi_\text{CDW} = L^{2-\eta} f[L^z/(T-T^\text{CDW}_c)]\,,
\end{equation}
with $\eta=0.25$ known from the exact solution of the 2D classical Ising model.

\begin{figure*}
\includegraphics[width=\textwidth]{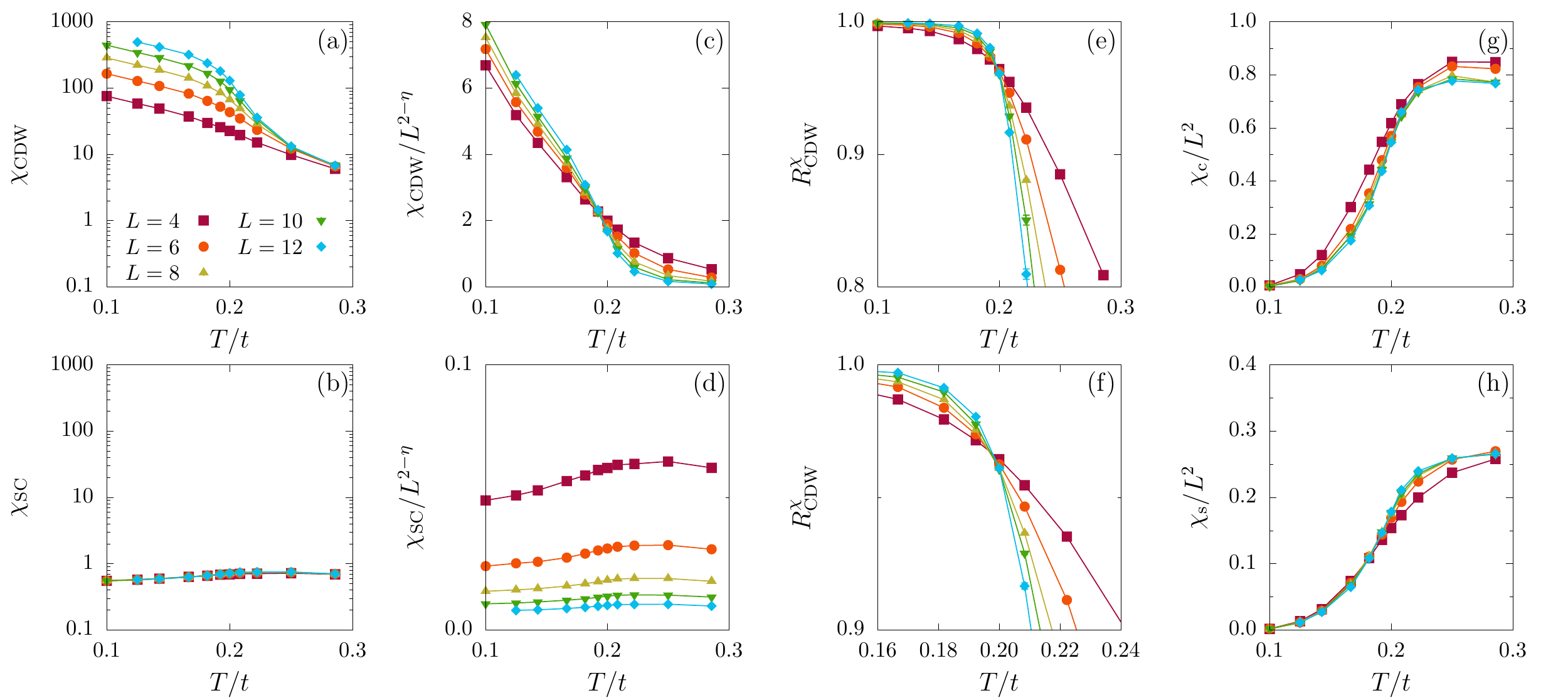}
\caption{\label{fig:om0.1_lambda0.075}
  Temperature-dependent (a) CDW susceptibility, (b) SC susceptibility, (c) rescaled CDW
  susceptibility ($\eta=0.25$), (d) rescaled SC
  susceptibility ($\eta=0.25$), (e)--(f) CDW correlation ratio [(f) shows a
  closeup], (g) uniform charge susceptibility, and (h) uniform spin
  susceptibility. Here, $\omega_0/t=0.1$, $\lambda=0.25$, and $L$ denotes the
  linear system size.}
\end{figure*}

\begin{figure*}
\includegraphics[width=\textwidth]{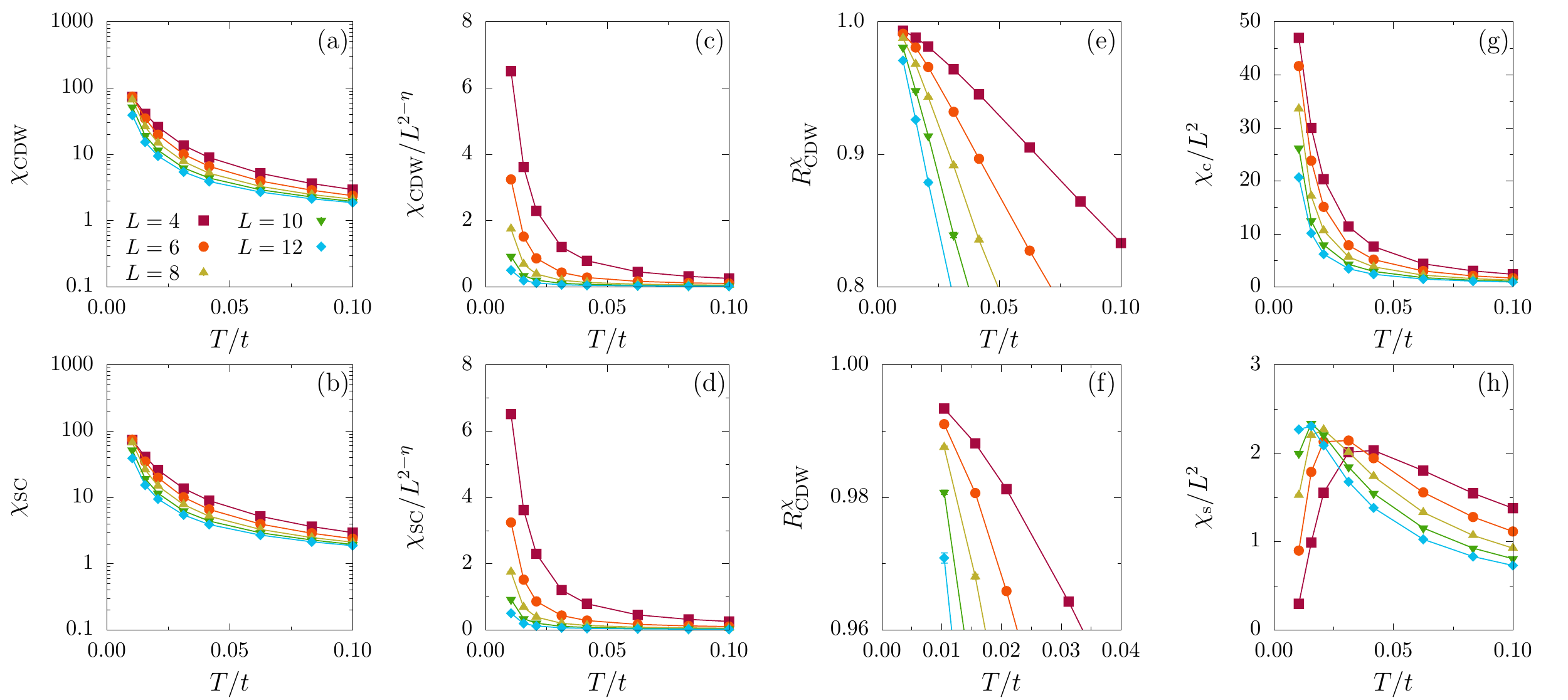}
\caption{\label{fig:ominf_lambda0.075}
  Same observables as in Fig.~\ref{fig:om0.1_lambda0.075} but for $\omega_0/t=\infty$
  and $\lambda=0.075$. These results were obtained directly from simulations of the attractive Hubbard model.}
\end{figure*}

A potential transition to an SC phase should be
in the Berezinskii-Kosterlitz-Thouless universality class with power-law
correlations below the critical temperature $T^\text{SC}_c$. In the absence of
long-range order and hence a divergence of $\chi_\text{SC}$, we exploit the
finite-size scaling form exactly at the critical temperature
\begin{equation}\label{eq:scalingSC}
  \chi_\text{SC} = L^{2-\eta}
\end{equation}
with $\eta=0.25$ \cite{kosterlitz1974critical}. Equation~(\ref{eq:scalingSC})
again implies a crossing point of results for different $L$ at
$T=T^\text{SC}_c$, as recently observed for the half-filled Holstein model on
the frustrated triangular lattice \cite{li2018superconductivity}.

\begin{figure*}
\includegraphics[width=\textwidth]{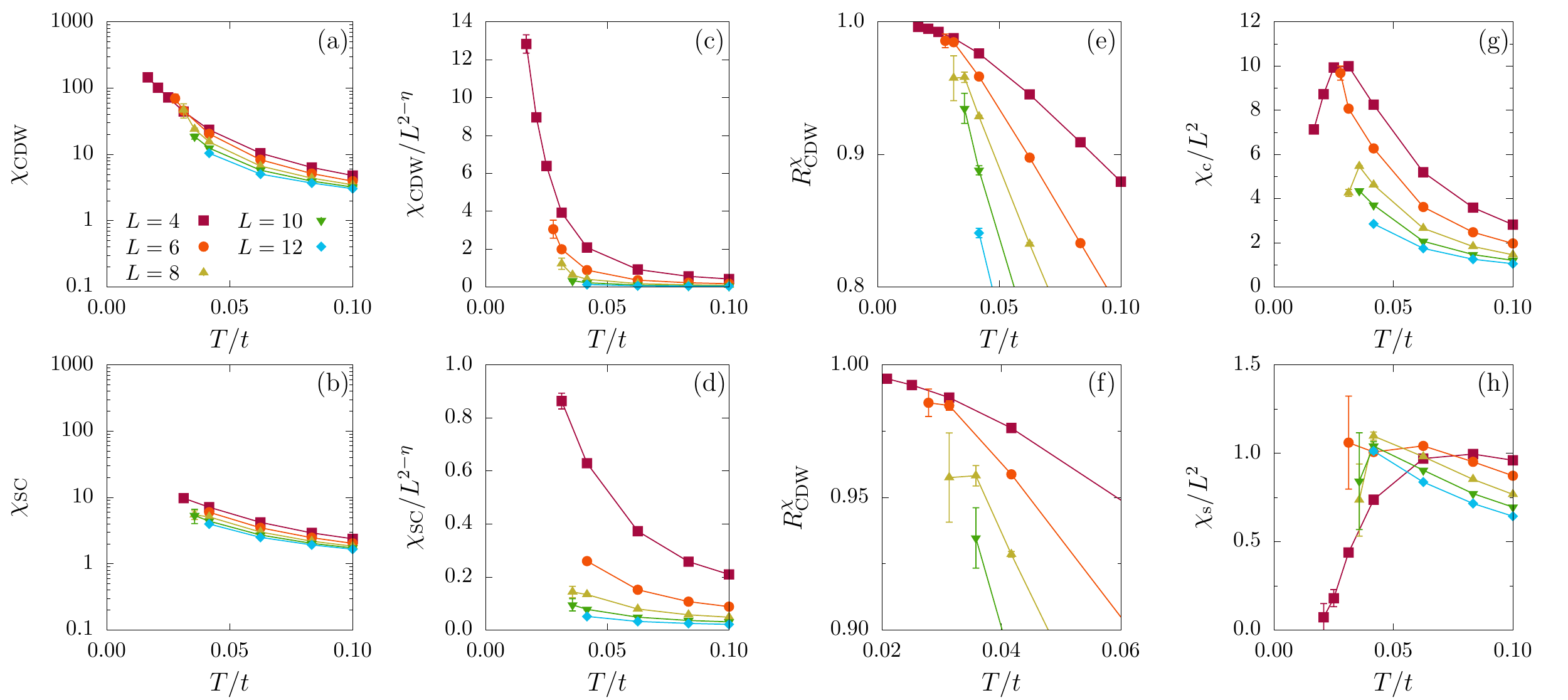}
\caption{\label{fig:om1.0_lambda0.075}
  As in Fig.~\ref{fig:om0.1_lambda0.075} but for $\omega_0/t=1$ and $\lambda=0.075$.}
\end{figure*}

\begin{figure*}
\includegraphics[width=\textwidth]{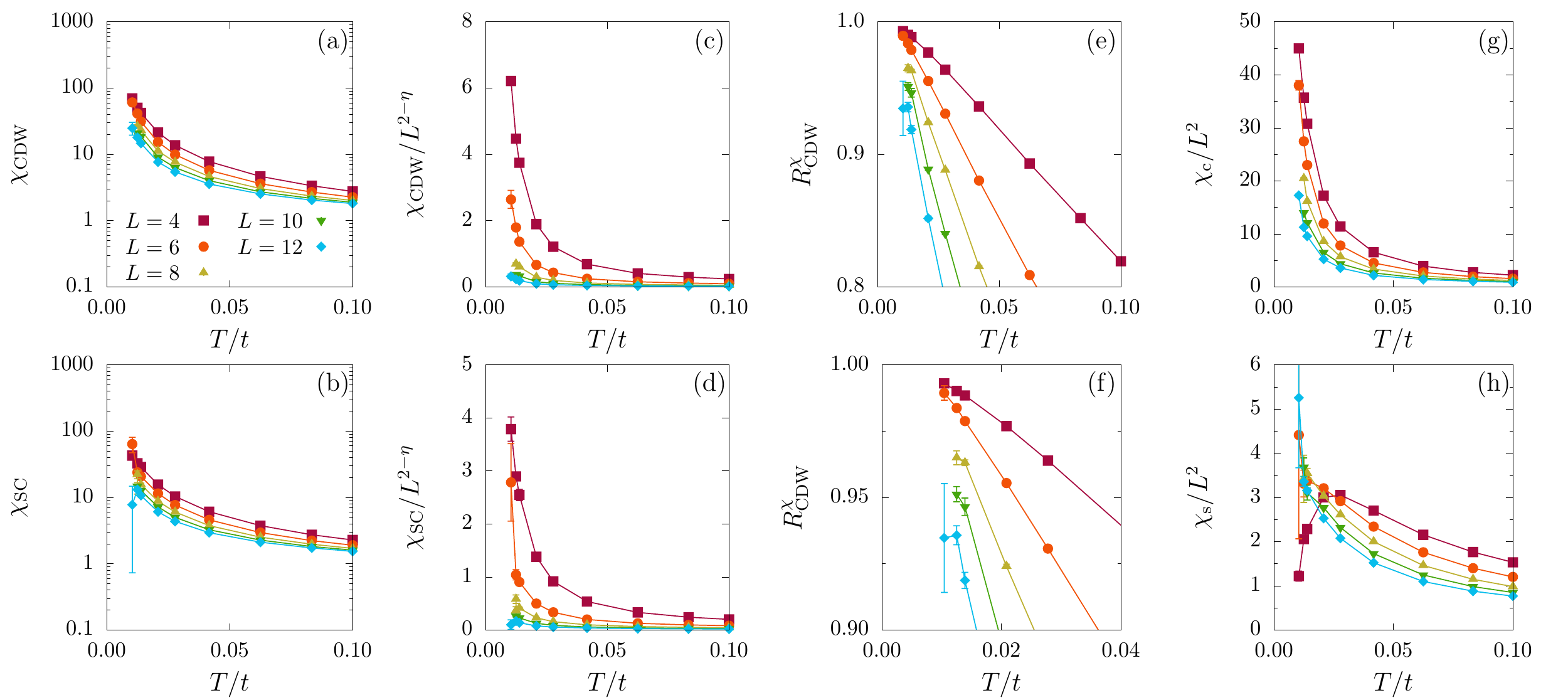}
\caption{\label{fig:om1.0_lambda0.025}
  As in Fig.~\ref{fig:om0.1_lambda0.075} but for $\omega_0/t=1$ and $\lambda=0.025$.}
\end{figure*}

\begin{figure}
\includegraphics[width=0.475\textwidth]{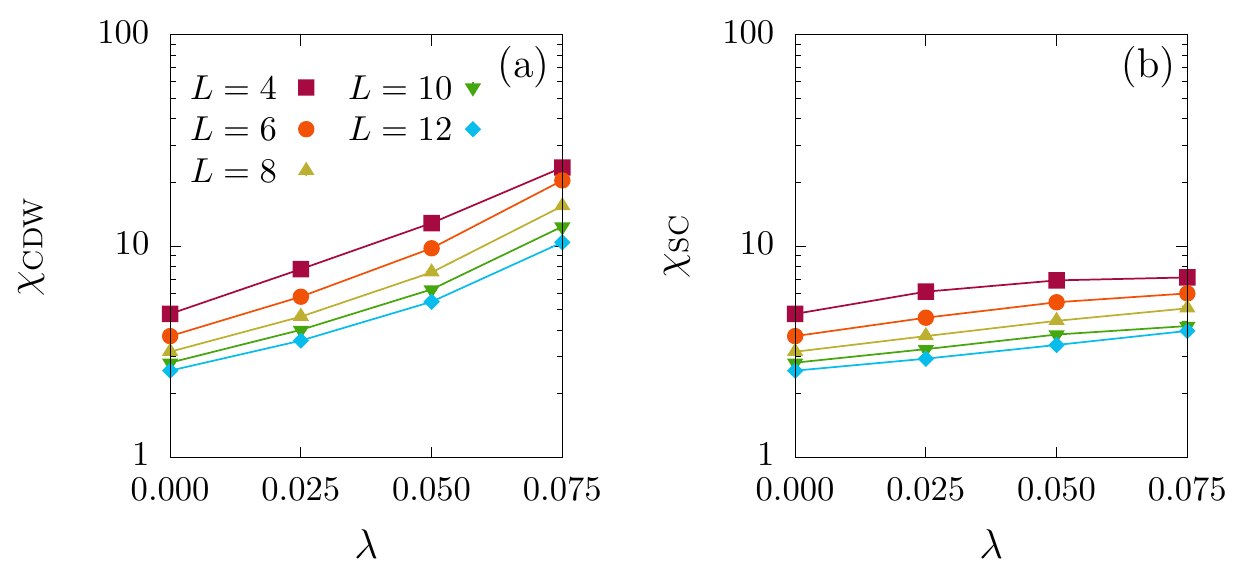}
\caption{\label{fig:om1.0_beta24}
(a) CDW and (b) SC susceptibilities as a function of $\lambda$ for $\omega_0/t=1$ and $T/t=1/24$.}
\end{figure}

\begin{figure}
\includegraphics[width=0.475\textwidth]{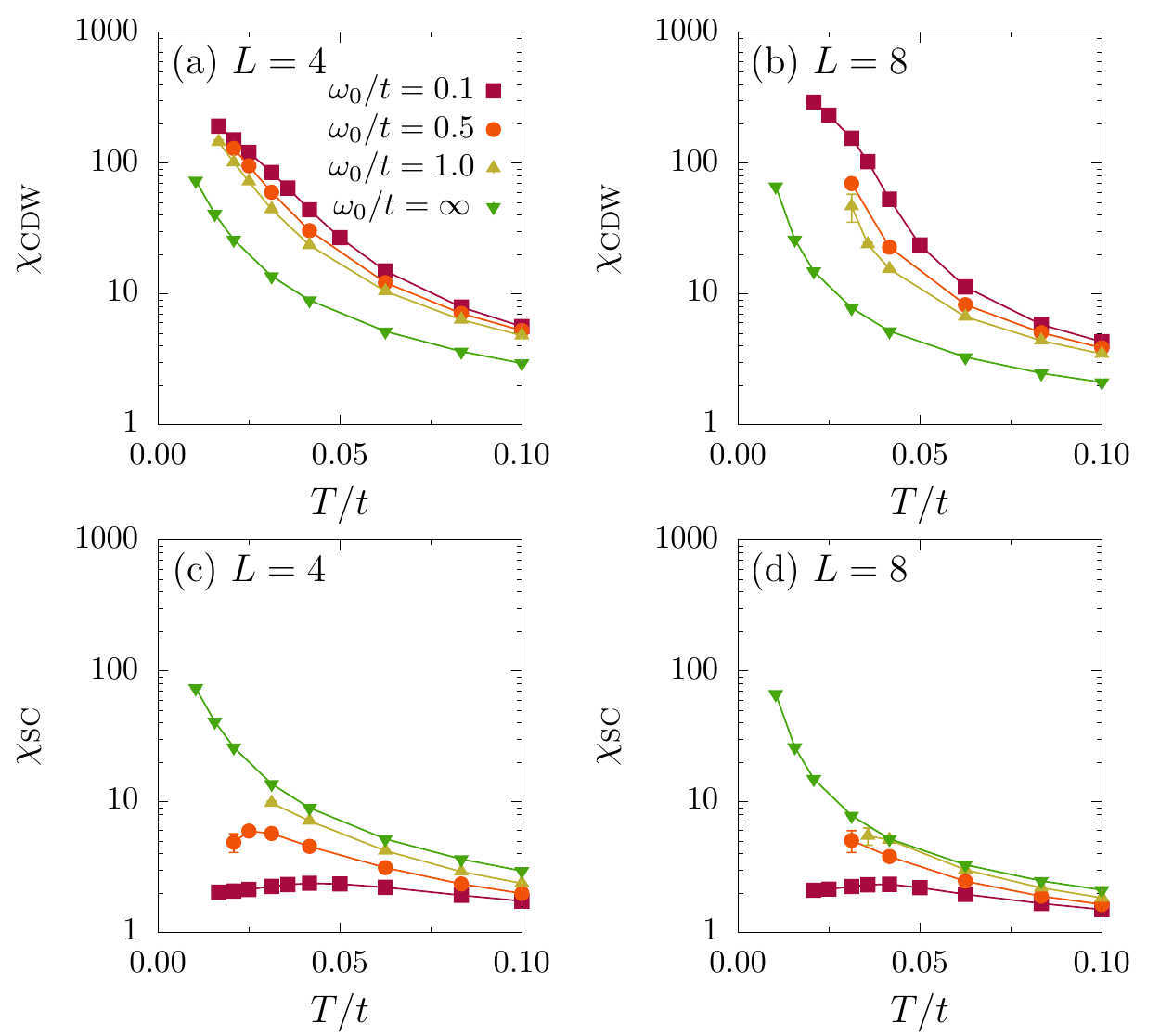}
\caption{\label{fig:lambda0.075_diffomgea}  
  CDW and SC susceptibilities for different $\omega_0/t$ and $L$. Here,
  $\lambda=0.075$. Results for $\omega_0=\infty$ were obtained directly
  from simulations of the attractive Hubbard model.}
\end{figure}

To detect gaps for long-wavelength charge and spin fluctuations, we also
consider the static (uniform) charge and spin susceptibilities 
\begin{align}\label{eq:localsusc}
\chi_\text{c} &=\beta \left(\las \hat{N}^2\ras - \las\hat{N}\ras^2\right)\,,&&\hspace*{-2em}\hat{N} = \sum_i \hat{n}_i\,, \\
\chi_\text{s} &=\beta \left(\las \hat{M}^2\ras - \las\hat{M}\ras^2\right)\,,&&\hspace*{-2em}\hat{M} = \sum_i \hat{S}^x_i\,. 
\end{align}
Here, $\hat{S}^x_i=\hat{c}^\dag_{i\UP}\hat{c}^\nag_{i\DO}+\hat{c}^\dag_{i\DO}\hat{c}^\nag_{i\UP}$,
  giving a maximum magnetization per site $\las M\ras/L^2=1$.

Based on the arguments in Sec.~\ref{sec:model}, we expect CDW order rather
than SC order at half-filling. Mean-field theory predicts a CDW transition
temperature $T^\text{CDW}_c\sim e^{-1/\sqrt{\lambda}}$ that appears consistent with
simulations for $\omega_0=0$ \cite{PhysRevB.98.085405} and renders the weak-coupling regime challenging.
Moreover, $T^\text{CDW}_c$ decreases with increasing $\omega_0$, and vanishes
for $\omega_0=\infty$ \cite{Hirsch85}. Before discussing the case of $\omega_0/t=1$
depicted in Fig.~\ref{fig:phasediagrams}, we consider $\omega_0/t=0.1$ (close
to the mean-field limit) and $\omega_0/t=\infty$ (the attractive Hubbard
model) as useful reference points. To address the findings of Refs.~\RefsAB
for $\omega_0/t=1$, we consider the couplings $\lambda=0.075$ and
$\lambda=0.025$, both inside the purported non-CDW phase in Fig.~\ref{fig:phasediagrams}.

Results for $\omega_0/t=0.1$ and  $\lambda=0.25$ are shown in
Fig.~\ref{fig:om0.1_lambda0.075}. The CDW transition for these parameters was
previously investigated using equal-time correlation functions
\cite{PhysRevB.98.085405}, so that we can benchmark our diagnostics.
 The data reveal a strong increase of the CDW
susceptibility at low temperatures [Fig.~\ref{fig:om0.1_lambda0.075}(a)],
whereas the SC susceptibility is virtually independent of $L$
[Fig.~\ref{fig:om0.1_lambda0.075}(b)]. The rescaled CDW susceptibility in
Fig.~\ref{fig:om0.1_lambda0.075}(c) exhibits a  clean crossing point
at $T^\text{CDW}_c/t\approx0.2$, in agreement with previous findings \cite{PhysRevB.98.085405}.
This crossing is consistent with that observed in the results for the
correlation ratio in Fig.~\ref{fig:om0.1_lambda0.075}(e), a close-up of which
is shown in Fig.~\ref{fig:om0.1_lambda0.075}(f). In contrast, the rescaled SC susceptibility
[Fig.~\ref{fig:om0.1_lambda0.075}(d)] is strongly suppressed for $T\lesssim T_c^\text{CDW}$. Finally, the uniform charge and spin
susceptibilities in Figs.~\ref{fig:om0.1_lambda0.075}(g) and (h),
respectively, reveal a gap in both sectors at sufficiently low temperatures,
as expected for a CDW insulator.

In the opposite, antiadiabatic regime $\omega_0=\infty$, we can rely on
previous results for the ground state of the attractive Hubbard model \cite{Hirsch85,Scalettar89}
to interpret our finite-temperature data. To gain insight into the
behavior expected in the weak-coupling regime, we consider $\lambda=0.075$
and focus on low temperatures $T/t\leq 0.1$.
The CT-INT data in Fig.~\ref{fig:ominf_lambda0.075} exhibit significant
differences compared to Fig.~\ref{fig:om0.1_lambda0.075}. The CDW and SC susceptibilities in
Fig.~\ref{fig:ominf_lambda0.075}(a) and Fig.~\ref{fig:ominf_lambda0.075}(b)
are identical due to the SO(4) symmetry of the Hubbard
model. Whereas a crossing point is not visible for the rescaled
susceptibilities in Figs.~\ref{fig:ominf_lambda0.075}(c) and (d) at the
temperatures considered, the CDW correlation ratio [Figs.~\ref{fig:ominf_lambda0.075}(e),(f)] again approaches 1 for $T\to 0$.
Such behavior is consistent with long-range CDW order at $T=0$.
Finally, the charge susceptibility in Fig.~\ref{fig:ominf_lambda0.075}(g) is
consistent with metallic behavior
(due to the coexistence of CDW and SC order), whereas
Fig.~\ref{fig:ominf_lambda0.075}(h) reveals the expected spin gap.

Having established the physics, but also the limitations of our simulations, in
the undisputed adiabatic and antiadiabatic limits, we turn to parameters that
are  directly relevant for Fig.~\ref{fig:phasediagrams}, specifically $\omega_0/t=1$ and
$\lambda=0.075$, where Refs.~\RefsAB predict a paramagnetic or SC state.
The corresponding results are shown in Fig.~\ref{fig:om1.0_lambda0.075}. Comparing the CDW and SC
susceptibilities in Figs.~\ref{fig:om1.0_lambda0.075}(a) and (b) reveals that
CDW correlations are significantly stronger than SC correlations at a given $T$.
Whereas a crossing point in the rescaled CDW susceptibility
[Fig.~\ref{fig:om1.0_lambda0.075}(c)] at temperatures below the accessible
range is plausible, we are unable to reach temperatures comparable to
$T_c^\text{CDW}$. On the other hand, a critical point signaling SC order is not expected based 
on the results of Fig.~\ref{fig:om1.0_lambda0.075}(d), especially upon
comparison with Fig.~\ref{fig:ominf_lambda0.075}(d) for
$\omega_0/t=\infty$. The latter case has stronger pairing correlations than
observed for $\omega_0/t=1$ even though $T^\text{SC}_c=0$. The CDW correlation ratio [Figs.~\ref{fig:om1.0_lambda0.075}(e),(f)] is also
consistent with CDW order at $T=0$.  The
uniform susceptibilities in Figs.~\ref{fig:om1.0_lambda0.075}(g) and (h) are
not entirely conclusive but consistent with a gap for charge and spin
excitations at $T=0$.

We also simulated a weaker coupling $\lambda=0.025$, deep inside the
predicted intermediate phase in Fig.~\ref{fig:phasediagrams}. Of course,
any type of order will be extremely delicate to
detect at such weak interactions on finite systems. Moreover, CDW and SC
correlations are necessarily degenerate at $\lambda=0$ (free
fermions). Nevertheless, Fig.~\ref{fig:om1.0_lambda0.025} does indicate
somewhat stronger CDW than SC correlations, which again seems to contradict the
claims of Refs.~\RefsAB. At the same time, the expected spin gap is only visible
in Fig.~\ref{fig:om1.0_lambda0.025}(h) at the lowest temperatures, whereas
the expected charge gap is beyond the accessible temperature range in Fig.~\ref{fig:om1.0_lambda0.025}(g).

The dependence of the CDW and SC susceptibilities on the coupling strength
$\lambda$ at $T=1/24$ can more clearly be seen in
Fig.~\ref{fig:om1.0_beta24}. Starting from identical values at $\lambda=0$,
$\chi_\text{CDW}$ increases significantly with $\lambda$,
whereas $\chi_\text{SC}$ flattens after a weak initial increase.

Finally, Fig.~\ref{fig:lambda0.075_diffomgea} compares the
temperature-dependent CDW and SC susceptibilities at different phonon
frequencies. The CDW susceptibility in Figs.~\ref{fig:lambda0.075_diffomgea}(a),(b)
evolves continuously, with values for
intermediate $\omega_0$ falling between those for $\omega_0/t=0.1$ and
$\omega_0/t=\infty$. For the SC susceptibility,
Figs.~\ref{fig:lambda0.075_diffomgea}(c),(d), the data suggest the
possibility of non-monotonic behavior: $\chi_\text{SC}$ for $\omega_0/t=1$ in
Fig.~\ref{fig:lambda0.075_diffomgea}(d) is equal to that for
$\omega_0/t=\infty$ at intermediate temperatures yet still smaller
than $\chi_\text{CDW}$.

\section{Conclusions}\label{sec:conclusions}

Although limitations regarding lattice size and temperature preclude
definitive conclusions regarding the ground state, we believe that our
unbiased results point rather strongly toward long-range CDW order in the
half-filled Holstein model on the square lattice. By extension, it seems
reasonable to expect only CDW and AFM ground states in the Holstein-Hubbard model.

Our main arguments are as follows.

(i) For the parameters considered, including those
where Refs.~\RefsAB predict no CDW order, we find that CDW
correlations are stronger than SC correlations, consistent with long-range
CDW order at $T=0$.

(ii) CDW (SC) correlations are stronger (weaker) than for the attractive
Hubbard model with the same effective interaction $U=\lambda W$. The latter corresponds to the
Holstein model in the antiadiabatic limit $\omega_0\to\infty$. Because the
Hubbard model is known to have long-range CDW order at $T=0$, this suggests
long-range CDW order also for the Holstein model with $\omega_0<\infty$. 
Weaker SC correlations do not  rule out SC order at $T=0$. However,
the coexistence of CDW and SC order in the attractive Hubbard model is
linked to an enhanced symmetry that is absent in the Holstein case for $\omega_0<\infty$
\cite{Hirsch83a}. Even if SC order exists at $T=0$, the stronger CDW order
conflicts with the claims of Refs.~\RefsAB.

(iii) Since we infer the nature of the ground state from simulations at $T>0$, there is in principle a possibility of a non-monotonic
temperature dependence, with a phase transition to an SC phase at even
lower temperatures. However, we do not observe any signatures or precursor
effects of this scenario, such as a decrease of the CDW susceptibility at
low temperatures.

(iv) Our results are consistent with the theoretical arguments for
a weak-coupling CDW instability due to nesting and a Van Hove singularity,
which should apply to the weak-coupling regime where a non-CDW region
was reported in Refs.~\RefsAB. 

It is beyond the scope of this work to determine the origin of the different
findings in Refs.~\RefsAB. However, the necessity of choosing
a variational wave function seems the most likely source for different
physics in trying to distinguish a paired Fermi liquid from either a
pair crystal (CDW state) or a pair condensate (SC state). In the
weak-coupling regime that is of interest here, different states are expected to be close in energy.
Moreover, the deviations between the critical values estimated with the help
of the same QMC method in Refs.~\RefsAB and visible in
Fig.~\ref{fig:phasediagrams}, also with respect to the strong-coupling phase
boundary $U=\lambda W$, suggest uncertainties that
significantly exceed the reported error bars. 

We expect the $T=0$ phase diagram in the ($\lambda$,$U$) plane to contain a single line of critical points that emanates from
the point $\lambda=U=0$ and separates CDW and AFM phases.
For further progress on this problem, functional RG calculations
with a suitable treatment of the energy and momentum dependence of the
interaction appear promising \cite{Barkim2015} to detect CDW order at weak coupling.
The combination of projective QMC simulations with improved updates based on
recent ideas \cite{BaSc2018,PhysRevB.98.041102,li2019accelerating} could
yield $T=0$ results without the variational approximation of Refs.~\RefsAB.
Finally, the use of pinning fields together with an extrapolation to the
thermodynamic limit may also prove useful \cite{Assaad13}.

\begin{acknowledgments}
We thank F. F. Assaad for helpful discussions and the DFG for
support via SFB 1170. We gratefully acknowledge the
computing time granted by the John von Neumann Institute for Computing (NIC)
and provided on the supercomputer JURECA \cite{jureca} at the J\"{u}lich
Supercomputing Centre.
\end{acknowledgments}

\end{document}